\documentclass[12pt]{article}

\usepackage{amssymb}
\usepackage[dvips]{graphicx}

\newcommand {\beq}{\begin{equation}}
\newcommand {\eeq}{\end{equation}}
\newcommand {\beqa}{\begin{eqnarray}}
\newcommand {\eeqa}{\end{eqnarray}}
\newcommand {\n}{\nonumber \\}
\newcommand {\tr}{\mbox{tr}}

\def\pa{\partial}
\renewcommand{\theequation}{\thesection.\arabic{equation}}

\begin{document}
\setlength{\oddsidemargin}{0cm}
\setlength{\baselineskip}{7mm}

\begin{titlepage}
\renewcommand{\thefootnote}{\fnsymbol{footnote}}
\begin{normalsize}
\begin{flushright}
\begin{tabular}{l}
OU-HET 496\\
October 2004
\end{tabular}
\end{flushright}
  \end{normalsize}

~~\\

\vspace*{0cm}
    \begin{Large}
       \begin{center}
         {M5-brane Effective Action as an On-shell Action in Supergravity}
       \end{center}
    \end{Large}
\vspace{1cm}

\begin{center}
           Matsuo S{\sc ato}$^{1)}$\footnote
            {
e-mail address : 
sato@pas.rochester.edu}
           {\sc and}
           Asato T{\sc suchiya}$^{2)}$\footnote
           {
e-mail address : tsuchiya@phys.sci.osaka-u.ac.jp}\\
      \vspace{1cm}
       
        $^{1)}$ {\it Department of Physics and Astronomy}\\
                {\it University of  Rochester, Rochester, NY 14627-0171, USA}\\
      \vspace{0.5cm}
                    
        $^{2)}$ {\it Department of Physics, Graduate School of  
                     Science}\\
               {\it Osaka University, Toyonaka, Osaka 560-0043, Japan}
\end{center}

\hspace{5cm}

\begin{abstract}
\noindent
We show that the covariant effective action for M5-brane is a solution
to the Hamilton-Jacobi (H-J) equations of 11-dimensional supergravity.
The solution to the H-J equations reproduces
the supergravity solution that represents the M2-M5 bound states. 
\end{abstract}
\vfill
\end{titlepage}
\vfil\eject

\setcounter{footnote}{0}

\section{Introduction}
\setcounter{equation}{0}
\renewcommand{\thefootnote}{\arabic{footnote}} 
D-branes and M-branes have been playing a crucial role in analyzing
nonperturbative aspects of string theory (M-theory).
D-branes can be regarded in string perturbation theory 
as a boundary state with the Dirichlet boundary condition imposed while they
also emerge as classical solutions of supergravity.

In a series of publications \cite{ST1,ST2,ST3}, we showed that
the D-brane effective action is a solution to the Hamilton-Jacobi (H-J)
equations of type IIA (IIB) supergravity 
and that the M2-brane and M5-brane effective actions are
solutions to the H-J equations of 11-dimensional supergravity.
We also showed that these solutions
to the H-J equations reproduce 
the supergravity solutions which represent a stack 
of D-branes in a $B_2$ field,
a stack of M2-branes and a stack of the M2-M5 bound states, respectively.
This fact means that those effective actions of branes
are on-shell actions around the corresponding 
brane solutions in supergravity. The near-horizon limit of these supergravity
solutions are conjectured to be dual 
to various gauge theories \cite{Maldacena,GKP-W}.
For instance, the near-horizon limit of the supergravity solution that
represents D3-branes in a $B_2$ field is conjectured to be dual to
noncommutative super Yang Mills in four dimensions \cite{HI,MR}. 
In gauge/gravity 
correspondence, the on-shell action
around the background dual to a gauge theory is in general a generating
functional of correlation functions in the gauge theory 
and the zero-mode part of the on-shell action should reproduce the holographic
renormalization flow \cite{dBVV}. Hence,
our findings should be useful for the study of the gauge/gravity (string) 
correspondence. (For other applications of our results, 
see the introduction in \cite{ST2}.)

In this paper, we revisit the M5-brane case.
The near-horizon limit of the supergravity solution representing the
M2-M5 bound states is conjectured to be dual to (a noncommutative version
of) 6-dimensional ${\cal N}=(2,0)$ superconformal 
gauge field theory \cite{Maldacena, MR}.
There are two versions \cite{M5action, covariantM5action} 
of the M5-brane effective action, which are
equivalent in the sense that both give the same equations of motion for
M5-brane \cite{eomforM5brane}.
One version of the action
was suggested in \cite{Witten} and explicitly constructed
in \cite{M5action}.
In this version, in order to obtain a complete set of the equations of
motion for M5-brane, one needs to add the self-duality condition
to the equations obtained by varying the action. In fact, we showed in 
\cite{ST2}
that this action is a solution to the H-J equations up to the self-duality
condition. This implies that this M5-brane effective action is not an on-shell
in the ordinary sense. So, it seems unclear what role this effective action as 
a solutions to the H-J equations plays in the gauge/gravity correspondence.
On the other hand, the other version of the action 
which was constructed
in \cite{covariantM5action} and is called the 'covariant action'
directly gives a complete set of the 
equations of motion for M5-brane. Although it contains an auxiliary scalar
field that causes a problem in defining 
the partition function \cite{Witten}, it is well-defined at least at classical
level. Thus, for the sake of the study of the 
gauge/gravity correspondence,
it is worth investigating whether this covariant action satisfies the
H-J equations of 11-dimensional supergravity and reproduces the supergravity
solution representing the M2-M5 bound states so that the action is an 
on-shell action around the M2-M5 bound state solution in the ordinary sense. 
In this paper, we 
show that this is indeed the case.

The present paper is organized as follows.
In section 2,  we perform a reduction of 11-dimensional supergravity
on $S^4$ and obtain a 7-dimensional gravity.
In section 3, we develop
the canonical formalism for the 7-dimensional gravity 
to derive the H-J equations. In section 4, we write down the covariant
action for M5-brane explicitly. In section 5, we show that the covariant
effective action is a solution to the H-J equations obtained in section 3.
In section 6, we show that the solution to the H-J equations in the previous
section reproduces the supergravity solution representing a stack of the M2-M5
bound states. In section 7, by using the relation of 11-dimensional 
supergravity
with type IIA supergravity, we obtain an effective action for  
NS 5-branes
that is a solution to the H-J equations of type IIA supergravity and
reproduces the supergravity solution representing a stack of 
NS 5-branes, which is 
relevant for the duality between gravity and little string theory \cite{ABKS}.
Section 8 is devoted to discussion. 
We give an argument
which is expected to intuitively explain
why the D-brane effective action satisfies the H-J equations of supergravity.
In appendix, we present some equations which are useful for the calculations
in sections 5 and 6.

\section{Reduction of 11-dimensional supergravity on $S^4$}
\setcounter{equation}{0}
In this section, we perform a reduction of 11-dimensional
supergravity on $S^4$ and
obtain a 7-dimensional gravity. 
We will regard a radial-time-fixed surface in the 7-dimensional gravity 
as a worldvolume of M5-brane.

The bosonic part of 11-dimensional supergravity is given by
\beqa
I_{11}=\frac{1}{2\kappa_{11}^{\:2}} \int d^{11}X \sqrt{-G}
\left(R_G-\frac{1}{2}|F_4|^2 \right) 
-\frac{1}{12\kappa_{11}^{\:2}} \int A_3 \wedge F_4 \wedge F_4,
\label{11Daction}
\eeqa
where
\beqa
F_4=d A_3
\label{fieldstrength}
\eeqa
and 
\beqa
|F_4|^2=\frac{1}{4!}F_{M_1M_2M_3M_4}F^{M_1M_2M_3M_4}.
\eeqa
We drop the fermionic degrees of freedom consistently.
The equations of motion derived from (\ref{11Daction}) are
\beqa
&&R^G_{MN}-\frac{1}{12}F_{M L_1L_2L_3}F_N^{\;\;L_1L_2L_3}
+G_{MN}\left(-\frac{1}{2}R_G+\frac{1}{4} |F_4|^2 \right)=0, \nonumber\\ 
&&D_L F^{L M_1 M_2 M_3} 
-\frac{1}{2 (4!)^2}\varepsilon^{M_1 M_2 M_3 L_1 \cdots L_8} 
F_{L_1 L_2 L_3 L_4} F_{L_5 L_6 L_7 L_8} = 0,
\label{EOM}
\eeqa
where $D_M$ stands for the covariant derivative in eleven dimensions, while
the Bianchi identity which follows from (\ref{fieldstrength}) is
\beqa
dF_4=0.
\label{Bianchi}
\eeqa

We split the 11-dimensional coordinates $X^M \;\; (M=0,1,\cdots,10)$ 
into two parts as $X^M=(\xi^\alpha,\theta_i) \;\;(\alpha=0,\cdots,6,\;\;
i=1,\cdots,4)$, where the $\xi^{\alpha}$ are 7-dimensional coordinates
and the $\theta_i$ parametrize $S^{4}$.
We make an ansatz for the fields as follows:
\beqa
ds_{11}&=&h_{\alpha\beta}(\xi)d\xi^{\alpha}d\xi^{\beta}
          +e^{\frac{\rho (\xi)}{2}} d\Omega_4, \n
F_4&=&\frac{1}{4!}F_{\alpha_1\cdots\alpha_4}(\xi)
             d\xi^{\alpha_1}\wedge\cdots\wedge d\xi^{\alpha_4} \n
          &&+\frac{1}{4!\:7!}e^{\rho(\xi)} \varepsilon^{\alpha_1\cdots\alpha_7}
             \tilde{F}_{\alpha_1\cdots\alpha_7}(\xi)
             \varepsilon_{\theta_{i_1}\cdots\theta_{i_4}}
             d\theta_{i_1}\wedge\cdots\wedge d\theta_{i_4}.
\label{ansatz}
\eeqa 
where $h_{\alpha\beta}$ is a 7-dimensional metric. 
Note that this ansatz preserves
the 7-dimensional general covariance.

By substituting (\ref{ansatz}) into the equations of motion (\ref{EOM}) and 
the Bianchi identity (\ref{Bianchi}), we obtain the following equations
in seven dimensions.
\beqa
&&R_{\alpha\beta}-\nabla_{\alpha}\nabla_{\beta}\rho
-\frac{1}{4}\pa_{\alpha}\rho\pa_{\beta}\rho
-\frac{1}{12}F_{\alpha\gamma_1\gamma_2\gamma_3}
F_{\beta}^{\;\;\gamma_1\gamma_2\gamma_3} \n
&&-\frac{1}{2}h_{\alpha\beta} \left( R+e^{-\frac{\rho}{2}}R^{(S^4)}
-2\nabla_{\gamma}\nabla^{\gamma}\rho 
-\frac{5}{4}\pa_{\gamma}\rho\pa^{\gamma}\rho
-\frac{1}{2}|F_4|^2+\frac{1}{2}|\tilde{F}_7|^2 \right)=0, \n
&&R+\frac{1}{2}e^{-\frac{1}{2}\rho}R^{(S^4)}
-\frac{3}{2}\nabla_{\gamma}\nabla^{\gamma}\rho 
-\frac{3}{4}\pa_{\gamma}\rho\pa^{\gamma}\rho
-\frac{1}{2}|F_4|^2-\frac{1}{2}|\tilde{F}_7|^2=0, \n
&&\nabla_{\gamma}(e^{\rho}F^{\gamma\alpha_1\alpha_2\alpha_3})
+\frac{1}{4!}e^{\rho}F_{\gamma_1\gamma_2\gamma_3\gamma_4}
\tilde{F}^{\alpha_1\alpha_2\alpha_3\gamma_1\gamma_2\gamma_3\gamma_4}=0, \n
&&\varepsilon^{\alpha\beta\gamma_1\gamma_2\gamma_3\gamma_4\gamma_5}
\pa_{\gamma_1}F_{\gamma_2\gamma_3\gamma_4\gamma_5}=0, \n
&&\nabla_{\gamma}(e^{\rho}
\tilde{F}^{\gamma\alpha_1\cdots\alpha_6})=0,
\eeqa
where $\nabla_{\alpha}$ stands for the covariant derivative 
in seven dimensions, and $R^{(S^4)}=12$.
By using a relation in seven dimensions, 
\beqa
\tilde{F}_{\alpha\gamma_1\cdots\gamma_6}
\tilde{F}_{\beta}^{\;\;\gamma_1\cdots\gamma_6}
=\frac{1}{7}h_{\alpha\beta}\tilde{F}_{\gamma_1\cdots\gamma_7}
\tilde{F}^{\gamma_1\cdots\gamma_7},
\nonumber
\eeqa
we can check that these equations are derived from the 7-dimensional
gravity given by
\beqa
I_7=\int d^7\xi \sqrt{-h} \: e^{\rho}
\left(R_h + e^{-\frac{\rho}{2}} R^{(S^4)}
+\frac{3}{4}\pa_{\alpha}\rho\:\pa^{\alpha}\rho
-\frac{1}{2}|F_4|^2 -\frac{1}{2}|\tilde{F}_7|^2 \right),
\eeqa
where 
\beqa
&&F_4=dA_3, \n
&&\tilde{F}_7=dA_6-\frac{1}{2}A_3\wedge F_4.
\eeqa
This reduction is a consistent truncation in the sense that 
every solution of $I_7$ can
be lifted to a solution of 11-dimensional supergravity.

\section{Canonical formalism and the H-J equations}
\setcounter{equation}{0}
In this section, we develop the canonical formalism for $I_7$
obtained in the previous section and derive the H-J equations.
First, we rename the 7-dimensional coordinates:
\beqa
\xi^{\mu}=x^{\mu} \;\; (\mu=0,\cdots,5), \;\;\; \xi^{6}=r.
\nonumber
\eeqa
Adopting $r$ as time, we make the ADM decomposition for the 7-dimensional
metric.
\beqa
ds_7^2&=&h_{\alpha\beta} \: d\xi^{\alpha}d\xi^{\beta} \n
&=&(n^2+g^{\mu\nu}n_{\mu}n_{\nu})\:dr^2+2n_{\mu}\:dr\:dx^{\mu}
+g_{\mu\nu}\:dx^{\mu}dx^{\nu},
\label{ADMdecomposition}
\eeqa
where $n$ and $n_{\mu}$ are the lapse function and the shift function,
respectively. Henceforce $\mu, \; \nu$ run from 0 to $5$.

In what follows, we consider a boundary surface 
specified by $r=\mbox{const.}$ and impose 
the Dirichlet condition for the fields on the boundary. Here we need to
add the Gibbons-Hawking term \cite{GH} to the actions, 
which is defined on the
boundary and ensures that the Dirichlet condition can be 
imposed consistently. 
Then, the 7-dimensional action $I_7$ with 
the Gibbons-Hawking term on the boundary can be expressed 
in the canonical form as follows:
\beqa
&&I_7 = \int dr d^6x 
\sqrt{-g}\:\left(\pi^{\mu\nu}\pa_r g_{\mu\nu}+\pi_{\rho}\pa_r \rho
+\pi^{\mu_1\mu_2\mu_3}\pa_r A_{\mu_1\mu_2\mu3}
+\pi^{\mu_1\cdots\mu_6}\pa_r A_{\mu_1\cdots\mu_6}\right.\n
&&\qquad \qquad \qquad \qquad  \;\;\;\left.
-nH-n_{\mu}H^{\mu}-A_{r\mu\nu}G_2^{\mu\nu}
-A_{r\mu_1\cdots\mu_5}G_5^{\mu_1\cdots\mu_5}\right)
\label{canonicalform}
\eeqa
with
\beqa
&&H=e^{-\rho}\left(-(\pi^{\mu\nu})^2+\frac{1}{9}(\pi^{\mu}_{\;\;\mu})^2
-\frac{5}{9}\pi_{\rho}^2+\frac{4}{9}\pi^{\mu}_{\;\;\mu}\pi_{\rho}
-3(\pi^{\mu_1\mu_2\mu_3}
+10\pi^{\mu_1\mu_2\mu_3\nu_1\nu_2\nu_3}A_{\nu_1\nu_2\nu_3})^2 \right. \n
&&\qquad\qquad\qquad \left.
-\frac{6!}{2}(\pi^{\mu_1\cdots\mu_6})^2\right)-{\cal L}, \n
&&H^{\mu}=-2\nabla^g_{\nu}\pi^{\mu\nu}+\pa^{\mu}\rho \pi_{\rho}
+F^{\mu}_{\;\;\nu_1\nu_2\nu_3}\pi^{\nu_1\nu_2\nu_3}
+\left(F^{\mu}_{\;\;\nu_1\cdots\nu_6}
-\frac{15}{2}A^{\mu}_{\;\;\nu_1\nu_2}F_{\nu_3\cdots\nu_6}\right)
\pi^{\nu_1\cdots\nu_6}, \n
&&G_2^{\mu\nu}=-3\nabla^g_{\lambda}\pi^{\lambda\mu\nu}, \n
&&G_5^{\mu_1\cdots\mu_5}=-6\nabla^g_{\rho}\pi^{\rho\mu_1\cdots\mu_5},
\label{HandHmuandG}
\eeqa
where
\beqa
{\cal L}=e^{\rho} \left(R_g-2\nabla^g_{\mu}\nabla_g^{\mu}\rho
-\frac{5}{4}\pa_{\mu}\rho\pa^{\mu}\rho
-\frac{1}{2}|F_4|^2-\frac{1}{2}|\tilde{F}_7|^2 \right)
+e^{\frac{\rho}{2}}R^{(S^{4})},
\label{L}
\eeqa
$\pi^{\mu\nu}$ is the canonical momentum conjugate to $g_{\mu\nu}$, and so on.
The relations between the canonical momenta and the $r$ derivatives of the
fields are given by
\beqa
&&\pi_{\mu\nu}=e^{\rho}\left(-K_{\mu\nu}+g_{\mu\nu}K
+\frac{1}{n}g_{\mu\nu}(\pa_r \rho -n^{\lambda}\pa_{\lambda}\rho)\right), \n
&&\pi_{\rho}=e^{\rho}\left(2K
+\frac{3}{2}\frac{1}{n}(\pa_r \rho-n^{\mu}\pa_{\mu}\rho)\right), \n
&&\pi_{\mu_1\mu_2\mu_3}=e^{\rho}\left(
-\frac{1}{6}\frac{1}{n}(F_{r\mu_1\mu_2\mu_3}-n^{\nu}F_{\nu\mu_1\mu_2\mu_3})
+\frac{1}{72}\frac{1}{n}
(\tilde{F}_{r\mu_1\mu_2\mu_3\nu_1\nu_2\nu_3}
-n^{\nu}\tilde{F}_{\nu\mu_1\mu_2\mu_3\nu_1\nu_2\nu_3})
A^{\nu_1\nu_2\nu_3}\right), \n
&&\pi_{\mu_1\cdots\mu_6}=-\frac{1}{6!}e^{\rho}\frac{1}{n}
(\tilde{F}_{r\mu_1\cdots\mu_6}-n^{\nu}\tilde{F}_{\nu\mu_1\cdots\mu_6}),
\label{canonicalmomenta}
\eeqa
where $K_{\mu\nu}$ is the extrinsic curvature given by
\beqa
K_{\mu\nu}=\frac{1}{2n}(\pa_r g_{\mu\nu}
-\nabla^g_{\mu}n_{\nu}-\nabla^g_{\nu}n_{\mu}), \;\;\; K=g^{\mu\nu}K_{\mu\nu}.
\eeqa
$n$, $n_{\mu}$ and $A_{r\mu\nu}$ and $A_{r\mu_1\cdots\mu_5}$ behave
like Lagrange multipliers and give the constraints:
\beqa
H=0,\;\;\; H^{\mu}=0,\;\;\; G_2^{\mu\nu}=0 \;\;\; \mbox{and} \;\;\; 
G_5^{\mu_1\cdots\mu_5}=0.
\label{constraints}
\eeqa
The first one and the second one are called the Hamiltonian constraint
and the momentum constraint respectively, while the third one and the last 
one are called the Gauss law constraints. Note that the hamiltonian density,
${\cal H}=nH+n_{\mu}H^{\mu}+A_{r\mu\nu}G_2^{\mu\nu}
+A_{r\mu_1\cdots\mu_5}G_5^{\mu_1\cdots\mu_5}$, vanishes on shell 
due to these constraints.

As usual, the H-J equation is obtained by performing the following 
replacements in the hamiltonian.
\beqa
&&\pi^{\mu\nu}(x)= 
\frac{1}{\sqrt{-g(x)}}\frac{\delta S}{\delta g_{\mu\nu}(x)}, \;\;\;
\pi_{\rho}(x)= 
\frac{1}{\sqrt{-g(x)}}\frac{\delta S}{\delta \rho(x)}, \;\;\;
\pi^{\mu_1\mu_2\mu_3}(x)= 
\frac{1}{\sqrt{-g(x)}}\frac{\delta S}{\delta A_{\mu_1\mu_2\mu_3}(x)}, \n
&&\pi^{\mu_1\cdots\mu_6}(x)= 
\frac{1}{\sqrt{-g(x)}}\frac{\delta S}{\delta A_{\mu_1\cdots\mu_6}(x)},
\label{replacements}
\eeqa
where $S$ is an on-shell action, and 
$g_{\mu\nu}(x)$, $\rho(x)$, $A_{\mu_1\mu_2\mu_3}(x)$ and 
$A_{\mu_1\cdots\mu_6}(x)$ represent the values of the fields on the
boundary $r=\mbox{const.}$.
The fact that the hamiltonian vanishes on shell simplifies 
the ordinary H-J equation:
\beqa
\frac{\pa S}{\pa r}+\int d^6x {\cal H} = 0 \; \rightarrow \;
\frac{\pa S}{\pa r}=0.
\label{ordinaryH-J}
\eeqa
This implies that $S$ does not depend on the boundary 'time' $r$
explicitly but depend only on the boundary values of the fields.
In addition to the ordinary H-J equation (\ref{ordinaryH-J}), there are
a set of equations for $S$ which is obtained by applying the replacements
(\ref{replacements}) to the constraints (\ref{constraints}). These equations
should also be called the H-J equations. For instance, the H-J equation
coming from $H^{\mu}=0$ takes the form
\beqa
&&-2\nabla^g_{\nu}
\left(\frac{1}{\sqrt{-g}}\frac{\delta S}{\delta g_{\mu\nu}}\right)
+\pa^{\mu}\rho \frac{1}{\sqrt{-g}}\frac{\delta S}{\delta \rho}
+F^{\mu}_{\;\;\nu_1\nu_2\nu_3}
\frac{1}{\sqrt{-g}}\frac{\delta S}{\delta A_{\nu_1\nu_2\nu_3}} \n
&&+\left(F^{\mu}_{\;\;\nu_1\cdots\nu_6}
-\frac{15}{2}A^{\mu}_{\;\;\nu_1\nu_2}F_{\nu_3\cdots\nu_6}\right)
\frac{1}{\sqrt{-g}}\frac{\delta S}{\delta A_{\nu_1\cdots\nu_6}}=0.
\eeqa
The H-J equations coming from $H^{\mu}=0$, $G_2^{\mu\nu}=0$ and
$G_5^{\mu_1\cdots\mu_5}=0$ gives a condition that 
$S$ must be invariant under the diffeomorphism in six dimensions
and the $U(1)$ gauge transformations
\beqa
&&A_3 \; \rightarrow \; A_3+d\Sigma_2, \n
&&A_6 \; \rightarrow \; A_6 +d\Sigma_5+\frac{1}{2}\Sigma_2 \wedge F_4.
\eeqa
(See appendix C in Ref.\cite{ST1}).
The H-J equation coming from $H=0$ is a nontrivial equation 
that can determine the form of
$S$. 

\section{Covariant M5-brane action}
\setcounter{equation}{0}
Before solving the H-J equations obtained in the previous section, we
write down the covariant effective action for M5-brane, which was
constructed in \cite{covariantM5action}.\footnote{For a canonical formulation
and a gauge fixing for the covariant M5-brane action, see Ref.\cite{DeCR}.}
It takes the form
\beqa
S_{M5}&=&-\int d^6\sigma  \left(\sqrt{
-\det({\cal G}_{\mu\nu}+\tilde{{\cal H}}_{\mu\nu})}
+\frac{1}{4}\sqrt{-{\cal G}}
\check{{\cal H}}^{\mu\nu}\tilde{{\cal H}}_{\mu\nu} \right) \n
&&+\int d^6\sigma \left( {\cal A}_6 +\frac{1}{2}{\cal A}_3 \wedge F_3 \right),
\label{M5action}
\eeqa
where the $\sigma^{\mu}$ ($\mu=0,\cdots,5$) parametrize the worldvolume
of the M5-brane and ${\cal G}_{\mu\nu}$, ${\cal A}_3$ and ${\cal A}_6$ are 
the induced fields on the worldvolume
of the corresponding fields in 11-dimensional supergravity. For instance, 
${\cal G}_{\mu\nu}$ is given by
\beqa
{\cal G}_{\mu\nu}(\sigma)=
\frac{\pa Y^M(\sigma)}{\pa\sigma^{\mu}}\frac{\pa Y^N(\sigma)}{\pa\sigma^{\nu}}
G_{MN}(Y(\sigma)),
\label{inducedfields}
\eeqa
where the $Y^M(\sigma)$ ($M=0,\cdots,10$) are embedding functions of the
worldvolume in eleven dimensions.
$F_3$ is the gauge field
strength on the worldvolume, which is a third-rank antisymmetric tensor.
There also exists an auxiliary scalar field on the worldvolume, which is 
denoted by $a$. It is convenient to introduce a time-like unit
vector field $v_{\mu}$:
\beqa
v_{\mu}=\frac{c_{\mu}}{\sqrt{-c_{\nu}c^{\nu}}}, \;\;\; 
c_{\mu}=\frac{\pa a}{\pa \sigma^{\mu}}, 
\;\;\;v_{\mu}v^{\mu}=-1.
\label{v}
\eeqa
Then $\check{\cal {H}}_{\mu\nu}$ and $\tilde{{\cal H}}_{\mu\nu}$ are defined
in terms of ${\cal G}_{\mu\nu}$, ${\cal A}_3$, $F_3$ and $v_{\mu}$ as follows:
\beqa
&&{\cal H}_{\mu\nu\lambda}={\cal A}_{\mu\nu\lambda}+F_{\mu\nu\lambda}, \n
&&{\cal H}^{*\mu\nu\lambda}
=\frac{1}{6}\varepsilon^{\mu\nu\lambda\rho\sigma\tau}
                      {\cal H}_{\rho\sigma\tau}, \n
&&\check{{\cal H}}_{\mu\nu}={\cal H}_{\mu\nu\lambda}v^{\lambda}, \n
&&\tilde{{\cal H}}_{\mu\nu}={\cal H}^{*}_{\mu\nu\lambda}v^{\lambda}.
\label{Hs}
\eeqa
Note that the effective action (\ref{M5action}) is the lowest order in
derivative expansion so that the effective action is valid only when
the fields are
almost independent of $\sigma$.

\section{Covariant M5-brane action as a solution to the H-J equations}
\setcounter{equation}{0}
In solving the H-J equations of the 7-dimensional gravity,
we drop the dependence of the fields
on the 6-dimensional coordinates $x^{\mu}$. Correspondingly, 
any solution of supergravity obtained from such a solution to the H-J 
equations will depend only on the radial time $r$. In other words, 
we reduce the problem to a one-dimensional one. 
Let $S_0$ be a solution to the H-J equations under this simplification. 
It follows from (\ref{HandHmuandG}), (\ref{L}) and
(\ref{replacements}) that the H-J equations coming from the hamiltonian
constraint $H=0$ is simplified as
\beqa
&&\left(\frac{1}{\sqrt{-g}}
\frac{\delta S_0}{\delta g_{\mu\nu}}\right)^2
-\frac{1}{9}\left(g_{\mu\nu}\frac{1}{\sqrt{-g}}
\frac{\delta S_0}{\delta g_{\mu\nu}}\right)^2
+\frac{5}{9}\left(\frac{1}{\sqrt{-g}} 
\frac{\delta S_0}{\delta \rho}\right)^2 \n
&&-\frac{4}{9}g_{\mu\nu}\frac{1}{\sqrt{-g}} 
\frac{\delta S_0}{\delta g_{\mu\nu}}
\frac{1}{\sqrt{-g}}\frac{\delta S_0}{\delta \rho} 
+3\left(\frac{1}{\sqrt{-g}}\frac{\delta S_0}{\delta A_{\mu\nu\lambda}}
+10A_{\rho_1\rho_2\rho_3}\frac{1}{\sqrt{-g}}
\frac{\delta S_0}{\delta A_{\mu\nu\lambda\rho_1\rho_2\rho_3}}\right)^2 \n
&&+\frac{6!}{2}
\left(\frac{1}{\sqrt{-g}}\frac{\delta S_0}{\delta A_{\mu_1\cdots\mu_6}}
\right)^2
+e^{\frac{3}{2}\rho}R^{(S^4)}=0.
\label{HJM5}
\eeqa
Let us consider the following form:
\beqa
S_0=S_c+S_{BI}+S_{WZ},
\label{S0}
\eeqa
with
\beqa
&&S_c=\alpha\int d^6 x \sqrt{-g}e^{\frac{3}{4}\rho}, \n
&&S_{BI}=\beta \int d^6 x \left(\sqrt{-\det(g_{\mu\nu}+\tilde{H}_{\mu\nu})}
+\frac{1}{4}\sqrt{-g}\check{H}^{\mu\nu}\tilde{H}_{\mu\nu} \right), \n
&&S_{WZ}=\gamma\int d^6 x \left( A_6 +\frac{1}{2}A_3 \wedge F_3 \right),
\label{S0s}
\eeqa
where $\check{H}_{\mu\nu}$ and $\tilde{H}_{\mu\nu}$ are defined in terms
of $g_{\mu\nu}$, $A_3$, $F_3$ and $v_{\mu}$ in the same way as (\ref{Hs}),
\beqa
&&H_{\mu\nu\lambda}=A_{\mu\nu\lambda}+F_{\mu\nu\lambda}, \n
&&H^{*\mu\nu\lambda}
=\frac{1}{6}\varepsilon^{\mu\nu\lambda\rho\sigma\tau}
                      H_{\rho\sigma\tau}, \n
&&\check{H}_{\mu\nu}=H_{\mu\nu\lambda}v^{\lambda}, \n
&&\tilde{H}_{\mu\nu}=H^{*}_{\mu\nu\lambda}v^{\lambda},
\label{checkHandtildeH}
\eeqa
and $v_{\mu}$ is defined in terms of $c_{\mu}$ in the same way as (\ref{v}).
All of the fields
in (\ref{S0s}) are independent of $x^{\mu}$ so that the integral over
the 6-dimensional space-time is factored out. In what follows, we
show that $S_0$ (\ref{S0}) is a solution to the simplified H-J equations 
of the 7-dimensional gravity with $c_{\mu}$ and $F_3$ being
arbitrary constants if
\beqa
\alpha^2=\frac{16}{3}R^{(S^4)}=64\;\;\; \mbox{and} \;\;\; \beta=-\gamma.
\label{condition}
\eeqa
$S_0$ trivially satisfies the simplified
H-J equations of the 7-dimensional gravity except (\ref{HJM5}),
while $S_0$ satisfies (\ref{HJM5}) quite nontrivially, as we will
see below.

If (\ref{S0}) is substituted into (\ref{HJM5}), the lefthand side of
(\ref{HJM5}) is decomposed into four parts:
\beqa
\mbox{LHS of (\ref{HJM5})}=(1)+(2)+(3)+(4)
\eeqa
with
\beqa
(1)&=&\left(\frac{1}{\sqrt{-g}}\frac{\delta S_c}{\delta g_{\mu\nu}}\right)^2
-\frac{1}{9}\left(g_{\mu\nu}
\frac{1}{\sqrt{-g}}\frac{\delta S_c}{\delta g_{\mu\nu}}\right)^2
+\frac{5}{9}\left(\frac{1}{\sqrt{-g}}\frac{\delta S_c}{\delta \rho}\right)^2
-\frac{4}{9}g_{\mu\nu}\frac{1}{\sqrt{-g}}\frac{\delta S_c}{\delta g_{\mu\nu}}
                      \frac{1}{\sqrt{-g}}\frac{\delta S_c}{\delta \rho} \n
&&+e^{\frac{3}{2}\rho}R^{(S^4)},\n
(2)&=&2g_{\mu\lambda}g_{\nu\rho}
\frac{1}{\sqrt{-g}}\frac{\delta S_c}{\delta g_{\mu\nu}}
\frac{1}{\sqrt{-g}}\frac{\delta S_{BI}}{\delta g_{\lambda\rho}}
-\frac{2}{9}g_{\mu\nu}g_{\lambda\rho}
\frac{1}{\sqrt{-g}}\frac{\delta S_c}{\delta g_{\mu\nu}}
\frac{1}{\sqrt{-g}}\frac{\delta S_{BI}}{\delta g_{\lambda\rho}}
-\frac{4}{9}g_{\mu\nu}
\frac{1}{\sqrt{-g}}\frac{\delta S_c}{\delta \rho}
\frac{1}{\sqrt{-g}}\frac{\delta S_{BI}}{\delta g_{\mu\nu}}, \n
(3)&=&\left(\frac{1}{\sqrt{-g}}\frac{\delta S_{BI}}{\delta g_{\mu\nu}}\right)^2
-\frac{1}{9}\left(g_{\mu\nu}
\frac{1}{\sqrt{-g}}\frac{\delta S_{BI}}{\delta g_{\mu\nu}}\right)^2 \n
&&+3\left(\frac{1}{\sqrt{-g}}\frac{\delta S_{BI}}{\delta A_{\mu\nu\lambda}}
+\frac{1}{\sqrt{-g}}\frac{\delta S_{WZ}}{\delta A_{\mu\nu\lambda}}
+10A_{\rho\sigma\tau}
\frac{1}{\sqrt{-g}}\frac{\delta S_{WZ}}{\delta A_{\mu\nu\lambda\rho\sigma\tau}}
\right)^2 ,\n
(4)&=&\frac{6!}{2}\left(
\frac{1}{\sqrt{-g}}\frac{\delta S_{WZ}}{\delta A_{\mu\nu\lambda\rho\sigma\tau}}
\right)^2.
\eeqa
By using the relations
\beqa
&&\frac{1}{\sqrt{-g}}\frac{\delta S_c}{\delta g_{\mu\nu}}
=\frac{1}{2}\alpha e^{\frac{3}{4}\rho} g^{\mu\nu}, \n
&&\frac{1}{\sqrt{-g}}\frac{\delta S_c}{\delta \rho}
=\frac{3}{4}\alpha e^{\frac{3}{4}\rho},
\label{derivativeofSc}
\eeqa
we can easily calculate (1), (2). The results are
\beqa
&&(1)=-\frac{3}{16}\alpha^2 e^{\frac{3}{2}\rho} 
+ e^{\frac{3}{2}\rho}R^{(S^4)}, \n
&&(2)=\left(1-\frac{2}{3}-\frac{1}{3}\right)g_{\mu\nu}
\frac{1}{\sqrt{-g}}\frac{\delta S_{BI}}{\delta g_{\mu\nu}}=0.
\eeqa
If the first condition in (\ref{condition}) holds, (1) also vanishes.
By using
\beqa
\frac{1}{\sqrt{-g}}\frac{\delta S_{WZ}}{\delta A_{\mu\nu\lambda\rho\sigma\tau}}
=\frac{\gamma}{6!}\varepsilon^{\mu\nu\lambda\rho\sigma\tau},
\label{derivativeofSWZ}
\eeqa
we can also calculate (4) easily:
\beqa
(4)=-\frac{\gamma^2}{2}.
\eeqa
The calculation of (3) is a nontrivial task. In appendix, we present 
some equations which are useful for the calculation of (3).
By using those equations and the second condition 
in (\ref{condition}), we obtain
\beqa
(3)=\frac{\gamma^2}{2}.
\eeqa
Thus (3) cancels (4). Therefore, (\ref{HJM5}) is actually satisfied with
$F_{\mu\nu\lambda}$ and $c_{\mu}$ being arbitrary constants.

In the remaining of this section, we show that our solution (\ref{S0}) can be
interpreted as the covariant M5 action (\ref{M5action}).
As is clear from (\ref{inducedfields}), if one sets in (\ref{M5action})
\beqa
Y^{\mu}(\sigma)=\sigma^{\mu}, \;\;\; \mbox{other }Y^M=0, \;\;\; 
\sigma^{\mu}=x^{\mu},
\eeqa
the induced fields in (\ref{M5action}) reduce to the fields in the 
7-dimensional gravity:
\beqa
{\cal G}_{\mu\nu}(\sigma)=g_{\mu\nu}(x),\;\;\; 
{\cal A}_{\mu\nu\lambda}(\sigma)=A_{\mu\nu\lambda}(x),\;\;\;
{\cal A}_{\mu_1\cdots\mu_6}(\sigma)=A_{\mu_1\cdots\mu_6}(x).
\eeqa
Hence, if $F_3$ and $c_{\mu}$ in (\ref{S0s}) are also 
identified with those in
(\ref{M5action}) and the $\sigma$-dependence of the fields in (\ref{M5action})
are completely neglected,
\beqa
S_{BI}+S_{WZ}=-\beta\:S_{M5}.
\eeqa
In other words, $S_{BI}+S_{WZ}$ is the effective action for M5-brane whose
worldvolume is not curved in the transverse directions. However, as 
in Ref.\cite{ST3}, we can show by taking into account 
the freedom of general coordinate
transformation that the effective action for M5-brane
whose worldvolume is curved in the $r$-direction is also a solution to
the H-J equations (up to a term analogous to $S_c$ in (\ref{S0})).
Whether the effective action for M5-brane with a general
configuration is a solution to the H-J equations of 11-dimensional supergravity
is an open problem.
Finally, we make a comment on $S_c$. We found in \cite{ST1,ST2} that the 
D-brane effective action with a term analogous to $S_c$ is a solution to 
the H-J equations of supergravity. The term analogous to $S_c$ has a dependence
on the dilaton field $\phi$ like $e^{-2\phi}$ while the D-brane effective
action has a dependence like $e^{-\phi}$ if the R-R fields are rescaled
appropriately. Therefore, the term analogous to $S_c$ should correspond 
to a contribution from the sphere amplitude in string theory. (Of course,
the D-brane effective action is a contribution from the disk amplitude.)
Thus, $S_c$ should correspond to a contribution from M-theory counterpart of 
the sphere amplitude in string theory.


\section{Supergravity solution of the M2-M5 bound state}
\setcounter{equation}{0}
The supergravity solution that represents a stack of $N$ M2-M5 bound states
is given in \cite{ILPT-RT}, 
and it is also a solution of $I_7$ which takes the following form:
\beqa
&&ds_7^2=f^{-\frac{2}{3}}k^{\frac{1}{3}}\eta_{\hat{\mu}\hat{\nu}}
dx^{\hat{\mu}}dx^{\hat{\mu}} 
+f^{\frac{1}{3}}k^{-\frac{2}{3}}\delta_{ab}dx^a dx^b 
+f^{\frac{1}{3}}k^{\frac{1}{3}}dr^2, \n
&&e^{\frac{\rho}{2}}=r^2 f^{\frac{1}{3}}k^{\frac{1}{3}} ,\;\;\;
A_{012}=\sin\theta f^{-1},\;\;\; A_{345}=\tan\theta k^{-1}, \n
&&\tilde{F}_{012345r}=3\cos\theta \tilde{Q} r^{-4} f^{-1} k^{-1},
\label{M2-M5}
\eeqa
where
\beqa
&&\hat{\mu},\;\hat{\nu}=0,1,2,\;\;\; a,\;b=3,4,5, \n
&&f=1+\frac{\tilde{Q}}{r^3},\;\;\; \tilde{Q}=\frac{\pi N}{\cos\theta},\;\;\;
k=\sin^2\theta+\cos^2\theta f,
\eeqa
$\theta$ is a parameter of the solution.
Note that this solution preserves sixteen supercharges and that
when $\theta=0$ the solution reduces to the ordinary solution representing
a stack of $N$ M5-branes.

Now that we obtained a solution (\ref{S0})
to the H-J equations in the previous section,
we have  a set of first-order differential equations, which can be regarded as
an integral of the equations of motion:
\beqa
\pi^{\mu\nu}=\frac{1}{\sqrt{-g}}\frac{\delta S_0}{\delta g_{\mu\nu}}, \;\;\;
\pi_{\rho}= \frac{1}{\sqrt{-g}}\frac{\delta S_0}{\delta \rho}, \;\;\;
\pi^{\mu\nu\lambda}=\frac{1}{\sqrt{-g}}
                      \frac{\delta S_0}{\delta A_{\mu\nu\lambda}}, \;\;\;
\pi^{\mu\nu\lambda\rho\sigma\tau}=\frac{1}{\sqrt{-g}}
\frac{\delta S_0}{\delta A_{\mu\nu\lambda\rho\sigma\tau}},
\label{1storderDE}
\eeqa
where (\ref{canonicalmomenta}) is substituted into the lefthand sides while
(\ref{S0}) into the righthand sides.

We evaluate the both sides of (\ref{1storderDE}) by substituting
the above solution (\ref{M2-M5}). 
The results for the lefthand sides are
\beqa
\pi_{\mu\nu}&=&\eta_{\mu\nu}\left(4r^3k^{\frac{5}{6}}f^{-\frac{1}{6}}
+\frac{1}{2}r^4k^{\frac{5}{6}}f^{-\frac{7}{6}}\pa_r f \right), \n
\pi_{ab}&=&\delta_{ab}\left( 4r^3f^{\frac{5}{6}}k^{-\frac{1}{6}}
+\frac{1}{2}r^4f^{\frac{5}{6}}k^{-\frac{7}{6}}\pa_r k \right), \n
\pi_{\rho}&=&6r^3 k^{\frac{1}{2}}f^{\frac{1}{2}}, \n
\pi_{012}&=&-\frac{1}{4}\sin\theta\tilde{Q}k^{\frac{1}{2}}f^{-\frac{3}{2}}, \n
\pi_{345}&=&-\frac{1}{4}\sin\theta\cos\theta
            \tilde{Q}k^{-\frac{3}{2}}f^{\frac{1}{2}}, \n
\pi_{012345}&=&-\frac{3}{6!}
               \cos\theta\tilde{Q}f^{-\frac{1}{2}}k^{-\frac{1}{2}},
\label{pisonM2-M5}
\eeqa
where the other $\pi_{\mu\nu\lambda}$ and 
$\pi_{\mu\nu\lambda\rho\sigma\tau}$ vanish.
We can also evaluate the righthand sides by using (\ref{derivativeofSc}),
(\ref{derivativeofSWZ}) and the equations in appendix.
The righthand sides coincide with the lefthand sides
(\ref{pisonM2-M5}) if we set in $S_0$
\beqa
&&F_{\mu\nu\lambda}=0, \n
&&\alpha=8,\n
&&\beta=-\gamma=-3\tilde{Q}\cos\theta.
\label{alphaandbeta}
\eeqa
The second and third equations are consistent with (\ref{condition}).
Thus the supergravity solution (\ref{M2-M5}) satisfies the first-order
differential equations (\ref{1storderDE}) given by $S_0$ (\ref{S0}). 
Hence, $S_0$ reproduces the supergravity solution (\ref{M2-M5}).
That is, $S_0$ is an on-shell action
around the supergravity solution (\ref{M2-M5}). Note that we need not
put any restriction on $c_{\mu}$ in order for (\ref{M2-M5}) to
satisfy (\ref{1storderDE}).

\section{NS 5-brane in type IIA supergravity}
\setcounter{equation}{0}
In this section, using the relation between 11-d supergravity and type IIA
supergravity, we obtain a solution to
the H-J equation of type IIA supergravity that
reproduces the supergravity solution representing a stack of 
NS 5-branes. So, this solution should correspond to a type IIA
NS 5-brane effective
action (plus a $S_c$-like term).
First, we consider a reduction of 11-d supergravity on $S^3\times S^1$, 
which is different from the one done in section 2:
\beqa
ds_{11}^2&=&G_{MN}dX^M dX^N \\
         &=&h_{\alpha\beta}(\xi)d\xi^{\alpha}d\xi^{\beta}
            +e^{\frac{1}{2}\rho_2(\xi)}d\Omega_3
            +e^{\frac{1}{2}\rho_1(\xi)}(dX^{10})^2, \n
F_4&=&\frac{1}{4!}F_{\alpha_1\cdots\alpha_4}
      d\xi^{\alpha_1}\wedge\cdots\wedge d\xi^{\alpha_4} \n
&&+\frac{1}{3!\:7!}e^{\frac{1}{4}\rho_1(\xi)+\frac{3}{4}\rho_2(\xi)}
\varepsilon_{\alpha_1\cdots\alpha_7}\tilde{F}^{\alpha_1\cdots\alpha_7}
\varepsilon_{\theta_{i_1}\theta_{i_2}\theta_{i_3}}d\theta_{i_1}\wedge
d\theta_{i_2}\wedge d\theta_{i_3} \wedge dX^{10},
\eeqa
where $\alpha,\;\beta$ run from 0 to 6, and $X^{10}$ parametrizes $S^1$.
Then, we obtain as a consistent truncation a seven-dimensional gravity
\beqa
&&I_7'=\int d^7\xi \sqrt{-h} \: e^{\frac{1}{4}\rho_1+\frac{3}{4}\rho_2}
\left(R_h + e^{-\frac{\rho_2}{2}} R^{(S^3)}
+\frac{3}{8}\pa_{\alpha}\rho_1\:\pa^{\alpha}\rho_2
+\frac{3}{8}\pa_{\alpha}\rho_2\:\pa^{\alpha}\rho_2 \right.\n
&&\qquad\qquad\qquad\qquad\qquad\qquad\;\;\;\;\;\left.
-\frac{1}{2}|F_4|^2 -\frac{1}{2}|\tilde{F}_7|^2 \right),
\eeqa
where $F_4=dA_3$, $\tilde{F}_7=dA_6-\frac{1}{2}A_3\wedge F_4$ 
and $R^{(S^3)}=6$. 
Let us consider the form
\beqa
S_0'=S_c'+S_{BI}+S_{WZ},
\label{S0'}
\eeqa
with
\beqa
S_c'=\tilde{\alpha}\int d^6 x \sqrt{-g}
e^{\frac{1}{4}\rho_1+\frac{1}{2}\rho_2},
\eeqa
and $S_{BI}$ and $S_{WZ}$ the same in (\ref{S0s}). We can 
verify that (\ref{S0'}) satisfies the H-J equation of $I_7'$ under
the same simplification as section 5 if
\beqa
\tilde{\alpha}^2=6R^{(S^3)}=36, \;\;\;\mbox{and} \;\;\;
\beta=-\gamma.
\eeqa

Next, following the relation between 11-d supergravity 
and type IIA supergravity,
we define the fields in type IIA supergravity in terms of those in
11-d supergravity as follows.
\beqa
&&h_{\alpha\beta}=e^{-\frac{2}{3}\phi}k_{\alpha\beta}, \n
&&\rho_1=\frac{8}{3}\phi, \;\;\; \rho_2=\rho-\frac{4}{3}\phi, \n
&&A_3=-C_3, \;\;\; A_6=-B_6.
\eeqa
$I_7'$ is rewritten in terms of these new fields as 
\beqa
&&I_7'=\int d^7\xi \sqrt{-k} \left[e^{-2\phi+\frac{3}{4}\rho}\left(
R_k+e^{-\frac{1}{2}\rho}R^{(S^3)}
+4\pa_{\alpha}\phi \pa^{\alpha}\phi 
+\frac{3}{8}\pa_{\alpha}\rho \pa^{\alpha}\rho
-3\pa_{\alpha}\phi \pa^{\alpha}\rho \right) \right. \n
&&\qquad\qquad\qquad\qquad\qquad\left.
-\frac{1}{2}e^{\frac{3}{4}\rho}|F_4^{(IIA)}|^2
-\frac{1}{2}e^{2\phi+\frac{3}{4}\rho}|\tilde{H}_7|^2 \right],
\label{I7IIA}
\eeqa
where $F_4^{(IIA)}=dC_3$ and $\tilde{H}_7=dB_6+\frac{1}{2}C_3\wedge F_4$.
This action is actually given by a consistent truncation 
of type IIA supergravity in which the ansatz for the fields is
\beqa
&&ds_{10}^2=k_{\alpha\beta}(\xi)d\xi^{\alpha}d\xi^{\beta}
+e^{\frac{1}{2}\rho(\xi)}d\Omega_3, \n
&&\phi=\phi(\xi), \n
&&H_3=-\frac{1}{3!\:7!}e^{2\phi+\frac{3}{4}\rho}
\varepsilon_{\alpha_1\cdots\alpha_7}\tilde{H}^{\alpha_1\cdots\alpha_7}(\xi)
\varepsilon_{\theta_{i_1}\theta_{i_2}\theta_{i_3}}
d\theta_{i_1}\wedge d\theta_{i_2} \wedge d\theta_{i_3}, \n
&&F_4^{(IIA)}=\frac{1}{4!}F^{(IIA)}_{\alpha_1\cdots\alpha_4}(\xi)
d\xi^{\alpha_1}\wedge\cdots\wedge d\xi^{\alpha_4},
\eeqa
where $H_3$ and $F_4^{(IIA)}$ are the NS-NS anti-symmetric field strength
and the R-R field strength, respectively.
Thus, by rewriting (\ref{S0'}) in terms of the fields 
in (\ref{I7IIA}), we obtain a solution to the H-J equations of (\ref{I7IIA}):
\beqa
S_0^{(NS5)}=S_c^{(NS5)}+S_{BI}^{(NS5)}+S_{WZ}^{(NS5)}
\label{NS5action}
\eeqa
with
\beqa
&&S_c^{(NS5)}=\tilde{\alpha}\int d^6x \sqrt{-g}e^{-2\phi+\frac{1}{2}\rho}, \n
&&S_{BI}^{(NS5)}=\beta \int d^6x e^{-2\phi}
\left(\sqrt{-\det(g_{\mu\nu}+e^{\phi}\tilde{H}_{\mu\nu}^{(NS5)})}
+\frac{1}{4}\sqrt{-g}\check{H}^{(NS5)\mu\nu}\tilde{H}_{\mu\nu}^{(NS5)}
\right), \n
&&S_{WZ}^{(NS5)}=\beta \int \left(B_6+\frac{1}{2}C_3 \wedge F_3\right),
\eeqa
where $g_{\mu\nu}$ is the $(\mu,\nu)$ component of $k_{\alpha\beta}$, and
$\check{H}_{\mu\nu}^{(NS5)}$ and $\tilde{H}_{\mu\nu}^{(NS5)}$ are defined
in terms of $H_{\mu\nu\lambda}^{(NS5)}=-C_{\mu\nu\lambda}+F_{\mu\nu\lambda}$
and $c_{\mu}$ in the same way as (\ref{checkHandtildeH}).
This solution to the H-J equations obviously
reproduces the supergravity solution of
type IIA NS 5-brane and should correspond to the NS 5-brane effective 
action (plus $S_c^{(NS5)}$ term). In fact, up to $S_c^{(NS5)}$, 
the solution (\ref{NS5action}) 
coincides with $\beta$ times the effective action for type IIA NS 5-brane 
which is proposed in \cite{BNS} if in the effective action
the R-R 1-form and its partner
1-form on the worldvolume are put to zero.

\section{Discussion}
\setcounter{equation}{0}
In this paper, we showed that the covariant effective action for M5-brane
is an on-shell action around the 
solution in 11-dimensional supergravity that represents a stack of the M2-M5
bound states. Applying our result to the gauge/gravity correspondence
is a future work.

In Refs.\cite{ST1,ST2}, we showed that the same thing holds 
for the D-brane case.
That is, the D-brane effective action 
(the Born-Infeld action plus the Wess-Zumino action)
is an on-shell action around the solution in type IIA(IIB) 
supergravity representing a stack of D-branes.
The following argument is expected to intuitively explain
why the D-brane effective action satisfies the H-J equations of supergravity
and is an on-shell action of supergravity.
Suppose that there exists a string field theory for type IIA or IIB 
superstring. There are two limits that can be taken for the string
field theory. One is the low energy limit (the $\alpha'\rightarrow 0$ limit),
and the other is the classical limit, $g_s\rightarrow 0$, where $g_s$ is the
string coupling. The string field
theory would reduce to type IIA or IIB supergravity
in the $\alpha' \rightarrow 0$ limit while in $g_s\rightarrow 0$ limit
it would reproduce the results
of the lowest order in the string perturbation theory.
As the string field theory is a quantum theory of gravity, 
wave functions or transition amplitudes in the string field theory 
should satisfy the equations
analogous to the Wheeler-DeWitt (WDW) equations, which we call the WDW-like
equations, where we regard the radial direction as time.
In $\alpha'\rightarrow 0$, these equations reduce to 
the WDW equations 
of the corresponding supergravity. Moreover, in the classical limit
($g_s \rightarrow 0$ limit), the WDW equations reduce to 
the H-J equations in the corresponding supergravity, 
which we are concerned with. Let us reverse the order of
these two limits. That is, when the $g_s \rightarrow 0$ limit is first taken in
the string field theory, one obtains from the WDW-like equations 
the equations of string theory 
which are analogous to the H-J equations and contain all 
$\alpha'$ corrections. We call these equations the H-J-like equations.
Note that it is difficult to write down these the H-J-like equations
because the conformal or the light-cone gauges do not seem to fit deriving
the equations.
Next, these H-J-like equations reduce in the
$\alpha'\rightarrow 0$ limit to the H-J equations in supergravity.

Let us consider a transition amplitude between the vacuum and a state that
represents a stack of $N$ D-branes in the string field theory. 
The amplitude must satisfy the WDW-like equations and reduces in 
$g_s \rightarrow 0$ limit to the 
transition amplitude between the vacuum and the D-brane 
boundary state, which must
satisfy the H-J-like equations. 
Moreover, under a condition that the fields on the worldvolumes of the D-branes
vary slowly, the amplitude is represented by the D-brane effective
action (the Born-Infeld action plus the Wess-Zumino terms). 
That is, the D-brane effective action satisfies the H-J-like
equations under the above condition. A nontrivial thing is that our results
show that the D-brane effective action satisfies the H-J equations itself.
Probably supersymmetry make this nontrivial thing possible.

Finally, let us check whether it is consistent 
that the solution to the H-J equations of supergravity
actually represents the effective action for a stack of $N$ D-branes.
As in the M5-brane case (see (\ref{alphaandbeta})), the Born-Infeld action
plus the Wess-Zumino terms in the solution is proportional to $N$ when the
solution is matched to the supergravity solution representing a stack of
$N$ D-branes. This is consistent with the fact that the tension of 
the stack of D-branes is proportional to $N$. The reason why
the solution to the H-J equations does not possess the non-Abelian
property is as follows. Neglecting the dependence on the worldvolume
coordinates in solving
the H-J equations implies imposing
$D_{\mu}F_{\nu\lambda}=0$ due to the gauge invariance, where
$F_{\mu\nu}$ is the gauge field strength on the worldvolume,
so that from the Bianchi identity $[F_{\mu\nu},F_{\lambda\rho}]=0$. 
Therefore, we cannot see the non-Abelian part of the gauge field strength.

In the near future, we hope to refine this argument and 
completely understand the reason why the D-brane effective action is 
a solution to the H-J equations of supergravity.


\section*{Acknowledgements}
We would like to thank D. Sorokin and N. Ishibashi for drawing our attention
to Refs.\cite{covariantM5action} and A. Das and S. Rajeev for discussions. 
The work of M.S. is supported in part by the US DOE Grant DE-FG 02-91ER40685.
The work of A.T. is supported in part by Grant-in-Aid for Scientific
Research (No.16740144) from the Ministry of Education, Science and Culture.

\section*{Appendix : Useful equations}
\setcounter{equation}{0}
\renewcommand{\theequation}{A.\arabic{equation}}
In this appendix, we present some equations which are useful 
for the calculations of (3)
in section 5  and of the canonical momenta in section 6.
By using a relation $\det \tilde{H}_{\mu\nu}=0$, which follows from 
$\tilde{H}_{\mu\nu}v^{\nu}=0$, one obtains
\beqa
\det(g_{\mu\nu}+\tilde{H}_{\mu\nu})=g\left(1-\frac{1}{2}\tr\tilde{H}^2
+\frac{1}{8}(\tr\tilde{H}^2)^2-\frac{1}{4}\tr\tilde{H}^4 \right) \equiv gE.
\label{det}
\eeqa
It is convenient to introduce an antisymmetric tensor $D_{\mu\nu}$:
\beqa
D_{\mu\nu}\equiv -\sqrt{\frac{1}{E}} \left( 
(1-\frac{1}{2}\tr\tilde{H}^2)\tilde{H}_{\mu\nu}
+\tilde{H}_{\mu\lambda}\tilde{H}^{\lambda\rho}\tilde{H}_{\rho\nu}\right).
\eeqa
In terms of $E$ and $D_{\mu\nu}$, the derivatives of $S_{BI}$ and $S_{WZ}$
are expressed shortly:
\beqa
&&\frac{1}{\sqrt{-g}}\frac{\delta S_{BI}}{\delta g_{\mu\nu}}
=\beta \left[
-\frac{1}{4}v^{\mu}v^{\nu}\tilde{H}^{\lambda\sigma}\check{H}_{\lambda\sigma}
-\frac{1}{8}\tilde{H}_{\lambda\sigma}
(H^{\lambda\sigma\mu}v^{\nu}+H^{\lambda\sigma\nu}v^{\mu})
+\frac{1}{2}\sqrt{E}g^{\mu\nu} \right.\n
&&\qquad\qquad\qquad\;\;\;\;\left.
+\frac{1}{2}D^{\mu}_{\;\lambda}\tilde{H}^{\lambda\nu}
+\frac{1}{4}(g^{\mu\nu}+v^{\mu}v^{\nu})
D^{\lambda\sigma}\tilde{H}_{\lambda\sigma} \right], \n
&&\frac{1}{\sqrt{-g}}\frac{\delta S_{BI}}{\delta A_{\mu\nu\lambda}} =
\beta\left[
\frac{1}{24}\varepsilon^{\mu\nu\lambda\rho\sigma\tau}
(2D_{\rho\sigma}-\check{H}_{\rho\sigma})v_{\tau}
+\frac{1}{12}(\tilde{H}^{\mu\nu}v^{\lambda}+\tilde{H}^{\nu\lambda}v^{\mu}
+\tilde{H}^{\lambda\mu}v^{\nu}) \right], \n
&&\frac{1}{\sqrt{-g}}\frac{\delta S_{WZ}}{\delta A_{\mu\nu\lambda}}
=\frac{\gamma}{72}\varepsilon^{\mu\nu\lambda\rho\sigma\tau}
F_{\rho\sigma\tau}.
\label{derivatives}
\eeqa
(\ref{derivativeofSWZ}) and the last equation in (\ref{derivatives}) gives
\beqa
\frac{1}{\sqrt{-g}}\frac{\delta S_{WZ}}{\delta A_{\mu\nu\lambda}}
+10A_{\rho\sigma\tau}
\frac{1}{\sqrt{-g}}\frac{\delta S_{WZ}}{\delta A_{\mu\nu\lambda\rho\sigma\tau}}
=\frac{\gamma}{72}\varepsilon^{\mu\nu\lambda\rho\sigma\tau}H_{\rho\sigma\tau},
\eeqa
where the lefthand side is a combination appearing in (3).

In what follows, we evaluate the quantities needed in calculating the righthand
side of (\ref{1storderDE}). First, by substituting (\ref{M2-M5}), we obtain
\beqa
&&\tilde{H}_{01}=-\tan\theta k^{\frac{1}{6}}f^{-\frac{5}{6}}v_2, \;\;\;
\tilde{H}_{12}=\tan\theta k^{\frac{1}{6}}f^{-\frac{5}{6}}v_0, \;\;\;
\tilde{H}_{20}=-\tan\theta k^{\frac{1}{6}}f^{-\frac{5}{6}}v_1, \n
&&\tilde{H}_{34}=-\sin\theta k^{-\frac{5}{6}}f^{\frac{1}{6}}v_5, \;\;\;
\tilde{H}_{45}=-\sin\theta k^{-\frac{5}{6}}f^{\frac{1}{6}}v_3, \;\;\;
\tilde{H}_{53}=-\sin\theta k^{-\frac{5}{6}}f^{\frac{1}{6}}v_4, \;\;\; \n
&&\check{H}_{01}=\sin\theta k^{-\frac{1}{3}}f^{-\frac{1}{3}}v_2 ,\;\;\;
\check{H}_{12}=-\sin\theta k^{-\frac{1}{3}}f^{-\frac{1}{3}}v_0 ,\;\;\;
\check{H}_{20}=\sin\theta k^{-\frac{1}{3}}f^{-\frac{1}{3}}v_1 ,\;\;\; \n
&&\check{H}_{34}=\tan\theta k^{-\frac{1}{3}}f^{-\frac{1}{3}}v_5, \;\;\;
\check{H}_{45}=\tan\theta k^{-\frac{1}{3}}f^{-\frac{1}{3}}v_3, \;\;\;
\check{H}_{53}=\tan\theta k^{-\frac{1}{3}}f^{-\frac{1}{3}}v_4.
\label{valueofH}
\eeqa
Next, note that there is a relation between $v_{\mu}$, 
which follows from $v_{\mu}v^{\mu}=-1$,
\beqa
k^{-\frac{1}{3}}f^{\frac{2}{3}}t+k^{\frac{2}{3}}f^{-\frac{1}{3}}u=-1,
\eeqa
where
\beqa
t=-v_0^2+v_1^2+v_2^2,\;\;\; u=v_3^2+v_4^2+v_5^2.
\eeqa
By using this relation and (\ref{valueofH}), $E$ in (\ref{det}) 
is calculated as
\beqa
E=\frac{kf^{-1}}{\cos^2\theta}
(1+k^{-\frac{1}{3}}f^{-\frac{1}{3}}\sin^2\theta u)^2.
\eeqa
$\mbox{}$From this, we obtain
\beqa
D_{\mu\nu}=\check{H}_{\mu\nu} 
\eeqa
when (\ref{M2-M5}) is substituted.

\end{document}